\documentclass[aps,prc,showpacs,twocolumn,epsfig]{revtex4}
\usepackage[dvips]{graphicx}
\begin{document}

\draft
\title{Nuclear radius deduced from proton diffraction by a black nucleus}

\author{Akihisa Kohama,$^1$ Kei Iida,$^1$ and Kazuhiro Oyamatsu$^{1,2,3}$}
\affiliation{$^1$RIKEN (The Institute of Physical and Chemical Research),
2-1 Hirosawa, Wako-shi, Saitama 351-0198, Japan\\
$^2$Department of Media Theories and Production, Aichi Shukutoku
University, Nagakute, Nagakute-cho, Aichi-gun, Aichi 480-1197, Japan\\
$^3$Department of Physics, Nagoya University, Furo-cho, Chigusa-ku, 
Nagoya, Aichi 464-8602, Japan}

\date{\today}

\begin{abstract}
     We find a new method to deduce nuclear radii from proton-nucleus elastic 
scattering data.  In this method a nucleus is viewed as a ``black'' sphere.
A diffraction pattern of protons by this sphere is equivalent to that of 
the Fraunhofer diffraction
by a circular hole of the same radius embedded in a screen.
We determine the black sphere radius in such a way
as to reproduce the empirical value of the angle of the observed first 
diffraction peak.  It is useful to identify this radius multiplied by 
$\sqrt{3/5}$ with the root-mean-square matter radius of the target nucleus.
For most of stable isotopes of masses heavier than 50, 
it agrees, within the error bars, with the values that were deduced 
in previous elaborate analyses from the data obtained 
at proton incident energies higher than $\sim 800$ MeV.
\end{abstract}

\maketitle

     Size of atomic nuclei, one of the most fundamental nuclear properties, 
remains to be determined precisely.  
Most popularly, the size is deduced from
electron and proton elastic scattering off nuclei
\cite{Alk:PR,Cha:AP,Igo:RMP,Bat:ANP,Fro:MT}.  The charge 
radii are well determined due to our full understanding of the underlying
electromagnetic interactions \cite{Fro:MT,DeV:ATO,FB}, 
while deduction of the matter radii from the proton-nucleus scattering 
data depends on the scattering
theory, which is more or less approximate in the sense that the 
nucleon-nucleon interactions involved are not fully understood.  During the 
past three decades there have been many efforts of deducing the matter density
distributions, which are
based on various scattering theories incorporating empirical nucleon-nucleon 
scattering amplitudes, such as Glauber theory \cite{Igo:RMP,Alk:PR} and 
nonrelativistic and relativistic optical potential methods 
\cite{Ray:PR,Coop:PRC47,Clark:PRC67,Sakagu:PRC}.  
A systematic analysis of the data for 
a large number of nuclides, however, is still missing.   
In this paper we propose a 
method to deduce the root-mean-square (rms) matter radii, which
is powerful enough to allow us to perform such a systematic analysis. 
This method, in which we assume that the target nucleus is completely
absorptive to the incident proton and hence acts like a "black" sphere, 
is far simpler than the conventional methods. 
This approximation was originally used by Placzek and Bethe \cite{Bethe} 
in describing the elastic scattering of fast neutrons.

     The present method is useful for heavy stable nuclei for which the proton 
elastic scattering data are present, as we shall see.  
In the conventional framework to deduce the rms radius, 
one tries to reproduce empirical data for the differential 
cross section for scattering angles covering several diffraction maxima 
\cite{Alk:PR,Bat:ANP,jpsj}, whereas, in the present method, one has only to
analyze the data around a maximum in the small angle regime. 
Remarkably, these two methods turn out to be
similar in the deducibility of the radius.

     Elastic scattering data for more neutron-rich 
unstable nuclei are expected to be provided by radioactive ion beam 
facilities, such as GSI and Radioactive Ion Beam Factory in RIKEN.  
In a possible scheme, a beam of unstable 
nuclei, such as Ni and Sn isotopes, created in heavy-ion collisions is 
incident on proton targets, and the protons 
scattered therefrom are detected, 
leading to the measurement of differential elastic cross section. 
We expect the present method to be effective at deducing the rms radius
of unstable nuclei from such measurement.

      We begin by regarding a target nucleus for proton elastic scattering
as a black sphere of radius $a$.  This picture holds when the target 
nucleus is completely absorptive to the incident proton.  For high incident 
kinetic energy $T_p$ above $\sim800$ MeV, the optical potential for this 
reaction is strongly absorptive.  It can be 
essentially viewed as a superposition of the nucleon-nucleon scattering 
amplitude.  Since the imaginary part of the amplitude is dominant over the 
real part in this energy range \cite{Ray:PRC19,Ray:PRC20}, the black sphere 
picture is applicable to a first approximation.
We note that the black sphere picture is fairly successful in describing 
the elastic scattering of low energy $\alpha$ particles \cite{Bat:ANP,FBl}.  
It was also used for analyses of the scattering of intermediate-energy pions and 
low-energy antiprotons \cite{Bat:ANP}.

     Since one can regard the proton beam as a plane wave of momentum
$p_{\rm Lab}$ $=\sqrt{(T_p+m_p)^2-m_p^2}$ with the proton mass, $m_p$, 
the black sphere picture can be described in terms of wave optics.  
This picture reduces to a diffraction of the wave by a circular black 
disk of radius $a$ if the corresponding wave optics is close to the limit of 
geometrical optics, i.e., $a/\lambda_{\rm Lab} \gg 1$, 
where $\lambda_{\rm Lab}$ $= 2\pi/p_{\rm Lab}$ is the wave length. 
We will later consider the ranges of $T_p \gtrsim 800$ MeV and $A \gtrsim 50$,
for which $a/\lambda_{\rm Lab} \gg 1$ is satisfied.
According to Babinet's principle, this diffraction is in turn equivalent to 
the Fraunhofer diffraction by a hole of the same shape as the disk embedded 
in a screen \cite{Landau}.  The scattering
amplitude for this diffraction 
in the center-of-mass (c.m.) frame of the proton and the nucleus reads
\begin{equation}
    f({\bf q})=i p a J_1(qa)/q,
    \label{dif}
\end{equation}
where ${\bf q}$ is the momentum transfer, ${\bf p}$ is the proton momentum in 
the c.m.\ frame, and $J_n$ is the $n$-th order Bessel function.  We 
then obtain the differential cross section as 
$d\sigma/d\Omega$ $=|f({\bf q})|^2$.  
We remark that Eq.\ (\ref{dif}) can be obtained from the 
absorptive limit of the Glauber theory in which the phase shift function 
is approximated by  
$\exp[i \chi(b)] = \theta(b - a)$, where ${\bf b}$ is the impact parameter 
perpendicular to ${\bf p}$, and $f({\bf q})$ is given by \cite{Glau:Lec}
\begin{equation}
  f({\bf q}) = i p \int_0^\infty bdb \; J_0(q a) 
                \{1 - \exp[i \chi(b)] \}.
\end{equation}

     We assume that the density distribution of the black sphere is
uniform.  Then it is natural to introduce an rms black sphere radius, $r_{\rm BS}$, as
\begin{equation}
   r_{\rm BS}\equiv \sqrt{3/5}a. 
   \label{rbb}
\end{equation}
In this stage, we determine $a$ in such a way that the c.m.\ scattering angle
[$\theta_{\rm c.m.}\equiv2\sin^{-1}(q/2p)$] of the first maximum for the 
Fraunhofer diffraction agrees with that measured by proton-nucleus elastic 
scattering, $\theta_M$.  (Here we define the zeroth peak as that whose angle 
corresponds to $\theta_{\rm c.m.}=0$.)  We remark that the diffraction 
patterns for $\theta_{\rm c.m.} \gg \theta_M$
are distorted by multiple scattering effects \cite{Alk:PR}, 
which are beyond the scope of the black sphere picture.  
The radius, $a$, and the angle, $\theta_M$, are then related by
\begin{equation}
   2 p a \sin(\theta_M/2) = 5.1356 \cdots.
\end{equation}
Combining this with Eq.\ (\ref{rbb}), we may thus write
\begin{equation}
   r_{\rm BS} = \frac{3.9780\cdots}{2p\sin(\theta_M/2)}. 
   \label{rbb2}
\end{equation}
It is this formula that we use in the followings.  As we shall see for 
heavy stable nuclei, the values of $r_{\rm BS}$ that can be determined from
Eq.\ (\ref{rbb2}) agree well with the values of the rms matter radius,
$r_m$, deduced from elaborate scattering theories in previous works.
For the estimate of $r_{\rm BS}$ from the data for $T_p \gtrsim 800$ MeV and
$A \gtrsim 50$, one can use the approximate expression, 
\begin{equation}
    r_{\rm BS}\simeq 4.50\left(\frac{1~{\rm GeV}}{p_{\rm Lab}}\right)
                    \left(\frac{10~{\rm deg}}{\theta_M}\right)~{\rm fm}.
\end{equation}

     For $^{58}$Ni and $T_p=1047$ MeV, the diffraction pattern is 
calculated from Eq.\ (\ref{dif}) by setting the first peak angle 
at $\theta_M$.  The 
result is shown in Fig.\ 1 together with the experimental data \cite{Lom:NPA}.
The heights of the diffraction maxima and minima thus calculated deviate from 
the empirical values, because the black sphere picture does not allow for the 
surface diffuseness.  That is why we pay attention to
the scattering angles of the diffraction maxima and minima.  The relation of 
these scattering angles to nuclear sizes has been discussed by Amado {\it et 
al.}\ \cite{ADL} in the Glauber theory.  The diffraction maxima are more 
advantageous to our study 
than the diffraction minima.  This is partly because statistical uncertainties 
in the yield count are much smaller near the maxima than near the minima and 
partly because theoretical prediction of the minima is rather sensitive to 
spin-orbit and Coulomb interactions between nucleons \cite{Alk:PR}.  
We remark that the behavior around the first diffraction peak can be well 
reproduced by incorporating nucleon distributions similar to the rectangular 
distribution assumed here into the Glauber theory 
(see Fig.\ 1 in Ref.\ \cite{iioya}).

\begin{figure}[t]
\begin{center}
\includegraphics[width=7cm]{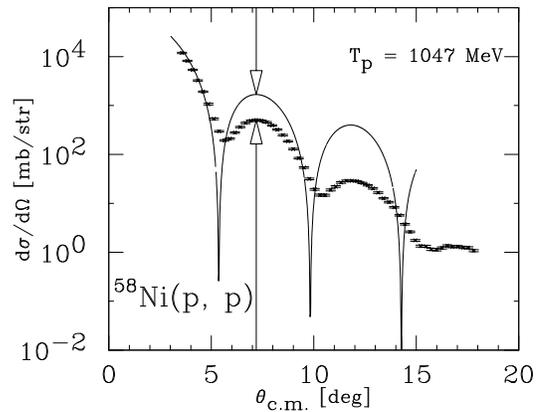}
\end{center}
\vspace{-0.5cm}
\caption{Differential cross section calculated from the Fraunhofer diffraction 
formula, Eq.\ (\ref{dif}), for $p$-$^{58}$Ni elastic scattering ($T_p=1047$ 
MeV).  The experimental data (crosses) are taken from Ref.\ \cite{Lom:NPA}.  
The arrows represent the first diffraction maximum, at which we fit the 
calculated peak angle to the empirical value.} 
\end{figure}

      It is important to note experimental uncertainty in the scattering
angle and in the proton incident energy 
since this gives rise to the uncertainty in the estimate of 
$r_{\rm BS}$, together with systematic errors that are dependent on the way of
determining the peak position.  The uncertainty in the measured angle, which is 
due mainly to the absolute angle calibration, is typically of order or 
smaller than $\pm0.03$ deg \cite{Ray:PRC18} for existing data for proton 
elastic scattering off stable nuclei,
while the uncertainty in the measured proton incident energy is typically
a few MeV \cite{Ray:PRC18}.

    In this work we focus on the proton elastic scattering data for $A\gtrsim
50$, $T_p \gtrsim 800$ MeV, and $\theta_{\rm c.m.}\gtrsim 5$ deg.  
We first display the Ni results for $r_{\rm BS}$, which are obtained from
the experimental values of $\theta_M$ using Eq.\ (\ref{rbb2}).   
We have used the data for $T_p = 796$ MeV \cite{Blan,Kyle,Hof:PLB},   
$T_p = 1040$ MeV \cite{Bertini}, and $T_p = 1047$ MeV \cite{Lom:NPA}. 
In collecting the data, we have made access to
Experimental Nuclear Reaction Data File (EXFOR [CSISRS]) \cite{exfor}. 
In Fig.\ 2, the values of $r_{\rm BS}$ derived from these data are plotted 
together with the deduced values of $r_m$. 
When the values of $r_m$ are not explicitly given in the literatures, we 
obtain them from $r_m^2= (Z/A) r_p^2 + [(A - Z)/A] r_n^2$, 
where $Z$ is the charge number, and $r_p$ and $r_n$ are the rms radii of
the proton and neutron distributions. 
The values of $\theta_M$ are determined from the scattering angle 
that gives the maximum value of the cross section among discrete data near 
the first diffraction maximum.  The error bars of $r_{\rm BS}$ in the plot,
which are of order or even greater than $\pm0.05$ fm, are determined from 
the half width between the neighboring angles measured. 
Uncertainties of $r_{\rm BS}$ associated with the 
absolute angle calibration and the measurement of $T_p$, which are smaller 
than $\pm0.03$ fm and $\pm0.02$ fm, respectively, 
are not taken into account.  
When the error of $r_m$ is not available, we evaluate it from 
the errors of $r_n$ and $r_p$ given in the literatures.

     We find from Fig.\ 2 that for all the isotopes, the values of 
$r_{\rm BS}$ agree with those of $r_m$ within the error bars 
except for an only case.  
This result is remarkable since there is no fitting parameter other 
than $a$ in the black sphere picture in contrast to the past elaborate analyses.
We note that it is premature to ask whether we have to regard $r_{\rm BS}$ 
as an rms radius of the point nucleon distribution or as an rms
matter radius folded with the nucleon form factor.  In fact, the 
present estimate of $r_{\rm BS}$ contains errors larger than the difference 
between these two radii, which is typically $\sim0.07$ fm.

\begin{figure}[t]
\begin{center}
\includegraphics[width=7cm]{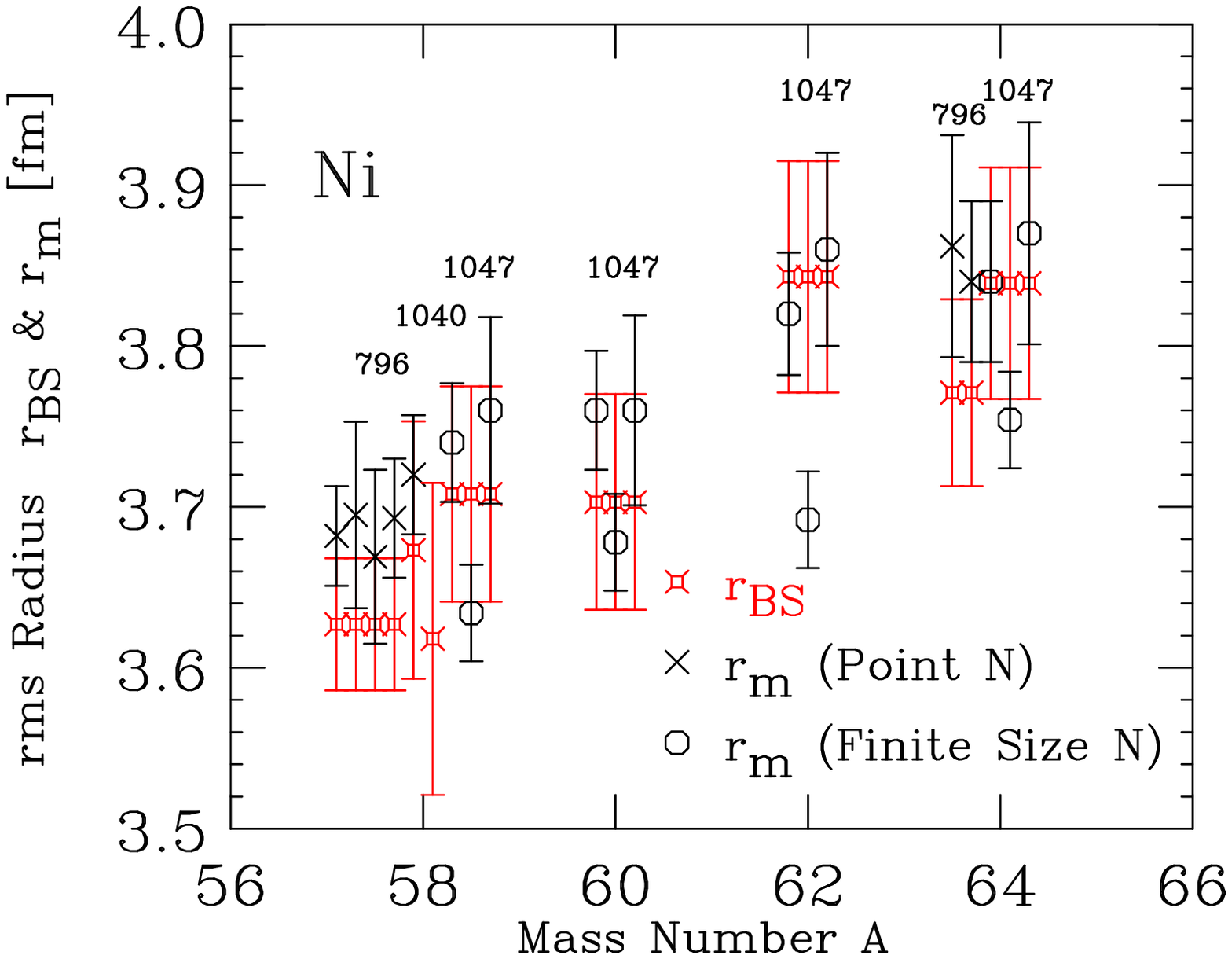}
\end{center}
\vspace{-0.5cm}
\caption{ (Color)
The values of $r_{\rm BS}$ (squares) for Ni isotopes.  
For the error bars, see text.  The values of $T_p$ (in MeV) are specified above these 
error bars.  For comparison we also plot the results for $r_m$ derived in the 
following references: For $A=58$, in Refs.\ \cite{Blan,Kyle,Ray:PRC18,Hof:PLB,Ray:PRC19} 
($T_p=796$ MeV) and in Refs.\ \cite{Alk:PLB,Chaum:PLB,Lom:NPA} ($T_p=1047$ 
MeV); for $A=60$ and 62, in Refs.\ 
\cite{Alk:PLB,Chaum:PLB,Lom:NPA}; for $A=64$, in Refs.\ \cite{Hof:PLB,Ray:PRC19} 
($T_p=796$ MeV) and in Refs.\ \cite{Alk:PLB,Chaum:PLB,Lom:NPA} ($T_p=1047$ 
MeV).  The crosses ($\times$) denote the rms matter radii of the point 
nucleon distributions, and the circles ($\circ$) denote those folded with the 
nucleon form factor.
Only $r_{\rm BS}$ is plotted for $A=58$ and $T_p=1040$ MeV, because no $r_m$ is 
available for this case. 
}
\end{figure}

      We next show the relations between $r_{\rm BS}$ and $r_m$ in Fig.\ 3, 
which are constructed from systematic data for various stable nuclides of mass
number larger than 50. In deriving $r_{\rm BS}$ 
we have used the data for $^{54}$Fe ($T_p=796$ MeV) \cite{Hof:PLB}, 
$^{90}$Zr ($T_p=800$ MeV) \cite{Hof:PRL}, 
$^{90}$Zr ($T_p=1000$ MeV) \cite{Alk:pre}, 
$^{116, 124}$Sn ($T_p=800$ MeV) \cite{Hof:PLB76},
and $^{208}$Pb ($T_p=800$ MeV) \cite{Blan,Hof:PRL}, 
$^{208}$Pb ($T_p=1000$ MeV) \cite{Alk:pre}, 
We do not include the data of $^{208}$Pb ($T_p = 1040$ MeV) \cite{Bertini} 
in this analysis, because the first peak position is not clear. 
The data for Ni isotopes are the same as used in Fig.\ 2.
We can see that the values of $r_{\rm BS}$ and 
$r_m$ including the error bars are mostly
on the line of $r_{\rm BS}=r_m$.  We thus
find that $r_{\rm BS}$ provides a good measure of the rms matter radius.  If 
elastic scattering data are obtained in a much finer manner, 
the main uncertainty in $r_{\rm BS}$ would 
arise from the absolute angle calibration 
and possibly the measurement of $T_p$.  In any case one could nicely 
determine the isotope dependence of $r_{\rm BS}$ if the relative peak angles 
between isotopes are accurately measured for the same proton beam.

   For nuclides for which elastic scattering data are available but 
no $r_m$, we list the values of $r_{\rm BS}$ for reference.  
For $^{90}$Zr ($T_p=800$ MeV) \cite{Gazza}, we obtain  
$r_{\rm BS}= 4.22 \pm 0.10$ fm; 
for $^{90}$Zr ($T_p=800$ MeV) \cite{Baker}, 
$r_{\rm BS}= 4.21 \pm 0.11$ fm;
for $^{92}$Zr ($T_p=800$ MeV) \cite{Baker},
$r_{\rm BS}= 4.21 \pm 0.07$ fm;
for $^{120}$Sn ($T_p=800$ MeV) \cite{Gazza},  $r_{\rm BS}$ $= 4.62 \pm 0.10$ fm;
for $^{144}$Sm ($T_p=800$ MeV) \cite{Gazza},  
$r_{\rm BS}= 4.92 \pm 0.27$ fm; 
for $^{154}$Sm ($T_p=800$ MeV) \cite{Barl}, 
$r_{\rm BS}= 5.24 \pm 0.09$ fm;
for $^{176}$Yb ($T_p=800$ MeV) \cite{Barl},
$r_{\rm BS}= 5.47 \pm 0.10$ fm;
for $^{208}$Pb ($T_p=800$ MeV) \cite{Gazza},  
$r_{\rm BS}= 5.54 \pm 0.29$ fm.
Among these nuclei, $^{154}$Sm and $^{176}$Yb are deformed in the ground state.
Since cross sections are generally measured for proton elastic scattering from randomly
oriented nuclei, the disk radius, $a$, obtained from the data lies between the
lengths of the semimajor and semiminor axes of the deformed nuclei.  As long as the 
degree of deformation is small, therefore, $r_{\rm BS}$ is expected to give a good 
measure of the rms matter radii of the deformed nuclei.

\begin{figure}[t]
\begin{center}
\includegraphics[width=7cm]{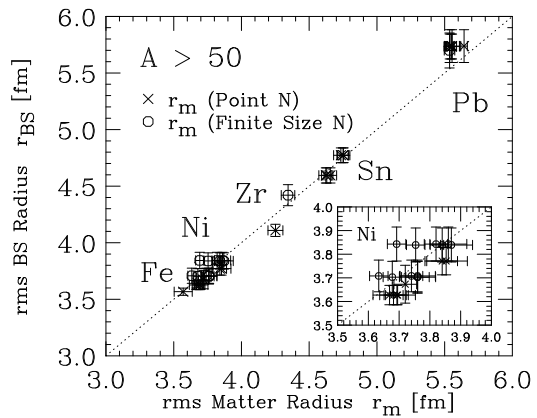}
\end{center}
\vspace{-0.5cm}
\caption{$r_{\rm BS}$ vs.\ $r_m$ for stable nuclei of masses above 50.
For Ni isotopes the values of $r_m$ are taken from the same references as 
cited in Fig.\ 2; for $^{54}$Fe, from Ref.\ \cite{Hof:PLB}; for $^{90}$Zr, 
from Refs.\ \cite{Ray:PRC18,Ray:PRC18a,Alk:NPA381}; for $^{116, 124}$Sn, from 
Refs.\ \cite{Ray:PRC19,Hof:PLB76,Ray:PRC18}; for $^{208}$Pb, 
from Refs.\ \cite{Ray:PRC19,Blan,Ray:PRC18,Ray:PRC18a,Blan:PRC,Hof:PRC21,Alk:NPA381}.  
The definition of the crosses ($\times$) and the circles ($\circ$) is
the same as in Fig.\ 2.  
The dotted line represents $r_{\rm BS}=r_m$.
Inset: $r_{\rm BS}$ vs.\ $r_m$ for Ni isotopes.
}
\end{figure}

     In summary we have performed a systematic analysis, based on the 
Fraunhofer diffraction off a black disk, of the existing data for proton 
elastic scattering off stable nuclei with masses larger than 50 at proton 
incident energies above $\sim800$ MeV.
This analysis allows us to make systematic estimates of the rms matter radii.  
The present method works even for the data in a range of ${\bf q}$ covering only 
the first diffraction maximum.  Such cases
would occur for neutron-rich unstable nuclei.  By combining the present result 
with a comprehensive table for the rms charge radii \cite{DeV:ATO,FB}, we can 
estimate neutron skin thickness for various nuclides on equal footing.  This 
is an important step towards understanding of the isospin-dependent bulk and 
surface properties of nuclear matter \cite{IO}.

       In the near future, an experiment that provides differential 
cross sections of proton elastic scattering off Ni unstable isotopes
will be performed in GSI.  In this experiment the projectile will be a 
radioactive ion beam having energy per nucleon of about 400 MeV, and
the scattering angles to be measured will contain the first diffraction 
maximum.  In applying the present prescription to estimate the rms matter 
radius to such measurements, its validity, which has been confirmed here
for proton incident energies above $\sim800$ MeV, needs to be examined at 
relatively low bombarding energies.   Investigation of how $r_{\rm BS}$ 
depends on the bombarding energy is now in progress \cite{KIO2}.  
Once the values of $r_{\rm BS}$ are accumulated for various 
nuclides including neutron-rich heavy nuclei, it is expected that
the density dependence of the symmetry energy near normal nuclear 
density will be clarified \cite{oyaii}.  
This expectation is also suggested by the work \cite{iioya} that pointed out 
the relation of $\theta_M$ with the density dependence of the symmetry energy.
We note, however, that the estimate of $r_{\rm BS}$ depends 
strongly on how sharp the energy distribution of a radioactive ion beam will 
be.

     In order to deduce the surface diffuseness in addition to the rms radius, 
one has to reproduce the overall behavior of the elastic scattering differential 
cross sections, as shown in Refs.\ \cite{Alk:PR,Bat:ANP,jpsj}.
The tails of the distributions, 
which are hard to deduce from the differential cross sections, might affect 
the prediction of reaction cross sections \cite{Ozawa}.  If reaction 
cross sections are measured for various heavy nuclides, 
the ``halo'' structure of unstable nuclei and 
the tails for stable nuclei will become clearer.

     We acknowledge K. Yazaki for his invaluable suggestions and comments. 
This work is supported in part by RIKEN Special Postdoctoral Researchers 
Grant No.\ A11-52040.

\end{document}